\documentclass[pdflatex,sn-basic,iicol]{sn-jnl}

\usepackage{graphicx}%
\usepackage{multirow}%
\usepackage{amsmath,amssymb,amsfonts}%
\usepackage{amsthm}%
\usepackage{mathrsfs}%
\usepackage[title]{appendix}%
\usepackage{xcolor}%
\usepackage{textcomp}%
\usepackage{manyfoot}%
\usepackage{booktabs}%
\usepackage{algorithm}%
\usepackage{algorithmicx}%
\usepackage{algpseudocode}%
\usepackage{listings}%

\begin{document}

\title[Phantom-Powered Nested Sampling]{Phantom-Powered Nested Sampling}

\author*[1]{\fnm{Joshua G.} \sur{Albert}}\email{albert@strw.leidenuniv.nl}

\affil*[1]{\orgdiv{Leiden Observatory}, \orgname{Leiden University}, \orgaddress{\street{PO Box 9513}, \city{Leiden}, \postcode{2300}, \country{The Netherlands}}}

\abstract{
We introduce a novel technique within the Nested Sampling framework to enhance efficiency of the computation of Bayesian evidence, a critical component in scientific data analysis. In higher dimensions, Nested Sampling relies on Markov Chain-based likelihood-constrained prior samplers, which generate numerous 'phantom points' during parameter space exploration. These points are too auto-correlated to be used in the standard Nested Sampling scheme and so are conventionally discarded, leading to waste. Our approach discovers a way to integrate these phantom points into the evidence calculation, thereby improving the efficiency of Nested Sampling without sacrificing accuracy. This is achieved by ensuring the points within the live set remain asymptotically i.i.d. uniformly distributed, allowing these points to contribute meaningfully to the final evidence estimation. We apply our method on several models, demonstrating substantial enhancements in sampling efficiency, that scales well in high-dimension. Our findings suggest that this approach can reduce the number of required likelihood evaluations by at least a factor of 5. This advancement holds considerable promise for improving the robustness and speed of statistical analyses over a wide range of fields, from astrophysics and cosmology to climate modelling.
}


\maketitle

\section{Introduction}\label{sec1}

Nested sampling \citep[NS;][]{2004AIPC..735..395S} is a Bayesian method of Lebesgue integration, enabling practical computation of the Bayesian evidence,
\begin{align}
    Z = \int_\mathcal{A} L(x) \,\mathrm{d}p(x),\label{eq:Z_lebesgue}
\end{align}
where $p(x)$ is a probability measure of a random variable, $x \in \mathcal{A}$, and $L(x)$ is a likelihood function. 
More generally, NS applies to scenarios where $L$ is any measurable function defined on a measurable space $(\mathcal{A}, \Sigma, p)$.
A naive quadrature approach to evaluating Eq.~\ref{eq:Z_lebesgue}, even when $L$ and $p$ are well-behaved, will quickly fail as the resolution of such a quadrature grid would quickly exceed hardware limits as the dimension of the problem grows. 
Nested sampling was introduced by John Skilling as a way to overcome this challenge, and does so by reformulating the problem of integration into a problem of compression.

Firstly, we define the likelihood constrained subset $\mathcal{A}(\lambda) = \{ x : L(x) > \lambda, x \in \mathcal{A}\}$, and its associated measure,
\begin{align}
    X_\lambda \triangleq p(\mathcal{A}(\lambda)) = \int_{\mathcal{A}(\lambda)}  \mathrm{d}p(x),\label{eq:X_lambda}
\end{align}
which we refer to as the enclosed prior volume.
Intuitively, one can see that as $\lambda$ increases the associated sub-space, $\mathcal{A}(\lambda)$, shrinks, and therefore $p(\mathcal{A}(\lambda))$ shrinks monotonically.
Now, consider any partitioning of likelihood values $(\lambda_0, \ldots, \lambda_n)$, where $\lambda_0$ is the infimum of likelihood values, and $\lambda_n$ is the supremum of likelihood values.
Associated with this partitioning is a sequence of sub-spaces $\mathcal{A}(\lambda_0) \subseteq \ldots \subseteq \mathcal{A}(\lambda_n)$.
Since the likelihood values partition the full range of the likelihood function, collectively the subsets cover the entire set,
\begin{align}
    \mathcal{A} =& \bigcup_{i=0}^n \mathcal{A}(\lambda_i)\\
    =& \bigcup_{i=1}^n \mathcal{A}(\lambda_i) \setminus \mathcal{A}(\lambda_{i-1}).
\end{align}
This last line writes the entire set in terms of the union of non-intersecting sets, and this then allows us to rewrite Eq.~\ref{eq:Z_lebesgue} as a sum,
\begin{align}
    Z = \lim_{n \to \infty} \sum_{i=1}^n \lambda_i (X_{\lambda_i} - X_{\lambda_{i-1}}).\label{eq:Z_X}
\end{align}

Equation~\ref{eq:Z_X} is incredible, as it has turned a difficult high-dimensional integral into a simple 1-dimensional sum.
The challenge and crux of NS then becomes in how to approximate the sequence $((\lambda_i, X_{\lambda_i}))_{i=0}^n$.
A natural Bayesian method for this falls from the construction above.

Starting from $i=0$ and $\lambda_0=0$ and $X_{\lambda_0}=1$ repeat the following:
\begin{enumerate}
    \item \textbf{Draw}: \textit{Uniformly} draw $x \sim p(\mathcal{A}(\lambda_i))$, and label $\lambda_{i+1}=L(x)$.
    \item \textbf{Shrink}: The corresponding subset measure will decrease by the order statistics of the \textit{uniform} distribution, $t \sim \mathcal{U}[0,1]$, so that $X_{\lambda_{i+1}} = t X_{\lambda_i}$.
    \item \textbf{Increment}: Stop if $\lambda_{i+1}$ is close to the supremum, otherwise $i \to i+1$.
\end{enumerate}

This is the so-called \textit{shrinkage} procedure, and is the general building block upon which all NS variations are built.
In the draw step, instead of drawing only a single sample, $x \sim p(\mathcal{A}(\lambda_i))$, a set of $m$ identically and independently uniformly distributed points can be drawn, and the smallest likelihood point selected, leading to a generalised order statistic $t \sim \mathrm{Beta}[m, 1]$.
In the shrink step, typically implementations keep track of expectations of $X_{\lambda}$, otherwise it can be Monte-Carlo simulated easily.
Finally, typically implementations have more nuanced stopping conditions than sensing if the likelihood supremum is reached, indeed, likelihoods need not be bounded.

Since its inception in \citet{2004AIPC..735..395S} numerous studies have improved all aspects of NS, from efficiently drawing from the likelihood-constrained prior \citep{2009MNRAS.398.1601F, 2011AIPC.1305..165B, 2015MNRAS.453.4384H, 2015EPJWC.10106019C, 2021PhRvD.103j3006W}, dynamically choosing the number, $m$, of points to draw \citep{2019S&C....29..891H}, parallel-computation \citep{2014AIPC.1636..100H}, and much more.
A wonderful review of the field is given by \citet{2023StSur..17..169B}.
Nested sampling has become an important tool in many fields, including cosmology  \citep{2006ApJ...638L..51M, 2007ApJ...660L..81E, 2009ApJ...699..985H, 2020A&A...641A..10P}, compact high-energy astrophysics \citep{2019ApJ...882L..24A, 2020arXiv200408342T}, spectra modelling \citep{2010MNRAS.401.2531A, 2014A&A...564A.125B}, gravitational lensing \citep{2010MNRAS.408.1969V, 2011MNRAS.415.2215B}, and more \citep{2011ApJ...729..106T,  2013ApJ...764..155L}.

The present paper introduces a methodology for vastly increasing the efficiency of nested sampling, i.e. decreasing the overall number of likelihood evaluations required to reach a target evidence accuracy. 
The paper is structured as follows: in Section~\ref{sec:formulation} we introduce a NS formulation that illuminates the opportunity for our proposed efficiency improvement; in Section~\ref{sec:method} we explain how we empirically validate our method, and in Section~\ref{sec:results} we show empirical results demonstrating the efficiency improvement; then in Section~\ref{sec:discussion} we discuss when this improvement can be employed, and its limitations.

\section{Formulating the opportunity for improving efficiency}
\label{sec:formulation}

One of the most challenging aspects of NS is sampling from the likelihood-constrained prior.
Many solutions to this problem maintain a set of simultaneously evolved samples within the likelihood constraint, which is called the `live set'.
For models of low dimension often rejection samplers, e.g. MultiNest \citep{2009MNRAS.398.1601F}, are the most efficient as there are enough samples in the live set at step $i$ to reconstruct convenient bounding representations of the subsets $\mathcal{A}(\lambda_i)$.
However, as model dimension grows, the samples will cluster near the outer shell of the subsets, leading to exponentially decaying sample efficiency. 
This is known as the curse of dimensionality.
Therefore, in higher dimensions Markov-chain samplers are used, e.g. PolyChord \citep{2015MNRAS.453.4384H}, which have polynomial decaying sample efficiency. 
Often these methods also require some analysis of the samples in the live set.

Markov-chain samplers work by creating a sequence of \textit{proposal} samples which satisfy the likelihood-constraint, with each sample conditionally dependent on the preceding sample.
As the number of steps increase the proposed samples become more independent from the initial point of the chain, and after a certain number of steps the chain can stop.
The last proposed sample becomes the \textit{accepted} sample.
The only necessary criterion for Markov-chain samplers is erogodicity, i.e. that eventually all possible points in the model space can be reached in a finite number of steps, and that the sampler cannot get stuck in periodic cycles.

The discarded samples, i.e. all-but-the-last proposed sample, are called the \textit{phantom} samples.
We note that discarding the phantom samples seems wasteful, as each proposal required hard work to generate a sample that satisfies the likelihood-constraint.
Therefore, we dig deeper into \textit{why} we discard phantom samples.

Recall that NS works by iteratively shrinking the enclosed prior volume, $X_\lambda$, and that this process depends on the order statistics of a set of $m$ likelihoods $\mathcal{S}=\{\lambda_i\}_{i=1}^m$.
The order statistic for shrinkage corresponding to the lowest likelihood is precisely described by $t \sim \mathrm{Beta}[m, 1]$ if all orderings of likelihoods have equal probability. 
If the orderings of likelihoods in $\mathcal{S}$ are not all equally likely, then the order statistics will deviate and the resulting evidence calculation will become biased.

So, the reason why phantom samples are discarded is to ensure the ordering of the resulting set $\mathcal{S}$ remains uniform. 
The question that then arises is, is discarding all phantom samples the best we can do, i.e. can we use some phantom samples and still keep $\mathcal{S}$ uniform?

\begin{figure}
    \centering
    \def\svgwidth{\columnwidth}
    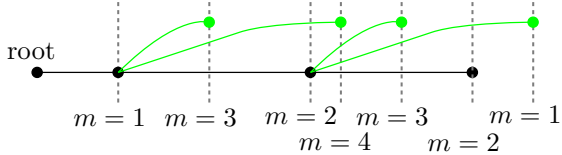
    \caption{A sample tree representations of samples produced from shrinkage process with a single Markov-chain, with phantom points in green and accepted samples in black, with horizontal position given by the likelihood of the sample. Edges connect samples to the likelihood constraint they were sampled within. Note, accepted samples connect only to other accepted samples. The size of $\mathcal{S}$, $m$, at each point is found by counting the number of edges crossed by vertical lines, shown by the grey dashed lines.}
    \label{fig:sample_tree}
\end{figure}

To illuminate the answer to this question we introduce a useful representation of the shrinkage process, which we call a \textit{sample tree}. 
A sample tree represents the samples produced by NS as a tree, where samples are nodes, and an edge connects each node to the node with likelihood label equal to the likelihood constraint that the sample was generated within.
The tree can then be drawn with horizontal position of each node equal to the likelihood label.
This is illustrated in Figure~\ref{fig:sample_tree} for a single Markov-chain.
A useful property of a sample tree, so arranged, is that $\mathcal{S}$ can be dynamically constructed for any likelihood constraint by drawing a vertical line and collecting all samples whose in-edges are crossed. 
Thus, the size of $\mathcal{S}$, $m$, for any likelihood constraint is given by counting edge crossings.
It is irrelevant if the sample tree is constructed by simultaneously evolving a set of `live points', or many parallel independent Markov-chains, or indeed if samples are added to it via any other means.
A similar representation was coincidentally introduced in \citet{2023StSur..17..169B}, however our sample tree was developed independently. 

A sample tree is useful for visualising how phantom samples are related to accepted samples.
An accepted sample and all its phantom samples (shown in green in Figure~\ref{fig:sample_tree}) share the same likelihood constraint, thus all have edges connected to the same \textit{sender node}.
Suppose there are $k$ phantom samples and $c$ parallel Markov-chains, as visualised Figure~\ref{fig:sample_tree_multi} in Appendix~\ref{app:proof}.
We make two observations.
Firstly, the last few phantom samples before the accepted sample are nearly identically sampled from the same distribution as the accepted sample, however they are not independent from one another.
This is because after enough proposal steps from an initial point, all further proposals can be considered uniformly sampled within the likelihood constraint, despite being correlated among themselves.
Secondly, the likelihoods in a sorted ordering of $\mathcal{S}$ will asymptotically approach uniformity as $c$ increases.
This is proven in Appendix~\ref{app:proof}. 

Based on these two observations we propose a method of boosting NS sampling efficiency when using a Markov-chain sampler:
\begin{enumerate}
    \item During each shrink step, take a large enough number of proposal steps, $i_{\rm min} + k$, so that samples from all steps $i > i_{\rm min}$ are identically sampled within the same likelihood constraint.
    \item Keep the last $k$ phantom samples, in addition to the accepted sample, by attaching them to your sample tree as shown in Figure~\ref{fig:sample_tree}.
    \item Choose a large enough number of parallel Markov-chain samplers $c$, so that the ordering of $\mathcal{S}$ approaches uniformity at all likelihoods. 
\end{enumerate}
This essentially says that, while it is crucial that all points in $\mathcal{S}$ are identically and independently uniformly drawn from within a likelihood constraint, we can ensure they are identically drawn, and asymptotically ensure they are independently drawn by following this simple procedure.

\section{Method}
\label{sec:method}

We empirically validate the method on a highly-correlated multivariate Gaussian likelihood with Gaussian prior model using the probabilistic programming framework JAXNS \citep{2020arXiv201215286A} to implement the model and the proposed method.
For the prior we consider a standard centred D-dimensional Gaussian of unit variance.
For the likelihood we consider a mean offset by 15 standard deviations from the prior, and a highly correlated covariance matrix with unit variance, and 99\% positive correlation between all dimensions. 
The evidence for this model is analytically known.
We evaluate the method over two model dimension sizes of $D=8$ and $D=16$.

For our Markov-chain likelihood-constrained sampler we use a standard 1-dimensional slice sampler \citep{2000physics...9028N}, with a few modifications. 
Firstly, in place of a step-out procedure, we initialise the bounding interval end points at the boundary of the model space.
While initially seeming naive, this avoids spending likelihood evaluations during the step out procedure, and perfectly satisfies detailed balance -- a stronger condition than ergodicity.
Secondly, during the contraction step, when a point along the slice is observed to fall outside the likelihood contour we shrink to the mid-point with the slice origin.
This increases the exponential contraction rate, while not impacting autocorrelation much \citep{10.1214/aos/1056562461}.
In sum, these two modifications result in a simple 1-dimensional slice sampler that quickly explores parameter space, while respecting the ergodicity requirement precisely.
It also lacks any additional analysis or meta-computation, e.g. clustering, required by some other implementations.

In order to probe the impact of Markov-chain autocorrelation, and determine $i_{\rm min}$, we explore the \textit{slice factor}, $s$, from which the number of proposal steps between accepted samples is $s D$.
Smaller values of $s$ will result in accepted sample samples that are less independent from the initial point of the chain, whereas higher values of $s$ will be more independent.
We explore values $s \in \mathcal{P}_s \triangleq  \{1,2,3,4,5,6\}$.

We retain the last $k$ phantom samples per accepted sample.
We explore values $k\in \mathcal{P}_k \triangleq  \{0,1,2,3,4,5\}$, where $k=0$ is standard nested sampling.

In order to study the convergence of the proposed method we explore number of parallel Markov-chains, $c \in \mathcal{P}_c \triangleq  \{10 D, 20 D, 30 D, 40 D\}$. 
In order to compare this with other standard nested sampling implementations this would be approximately equivalent to simultaneously evolving a set of $m=c (1 + k)$ live points. 

For each combination of $s$, $k$ and $c$ we run the nested sampling from 100 different random seeds. 
We compute ensemble averages of the log-evidence statistics, and all other run diagnostics.

Our primary aim is to measure the asymptotic convergence to standard nested sample, $k=0$.
Therefore, we define the target distribution as the run with $s_{\rm target}=\sup_s{\mathcal{P}_s}$, $c_{\rm target}=\sup_c{\mathcal{P}_c}$, and $k=0$.
Our primary measurement is bias from this target log-evidence distribution.

We note, there is a degeneracy in how to attribute likelihood evaluations \textit{per sample} for $k>0$.
In our implementations, we do this by counting the total number of likelihood evaluations per accepted sample and equally distributing them among the accepted sample and $k$ retained phantom samples.

\section{Results}
\label{sec:results}

\begin{figure*}
    \centering
    \includegraphics[width=2\columnwidth]{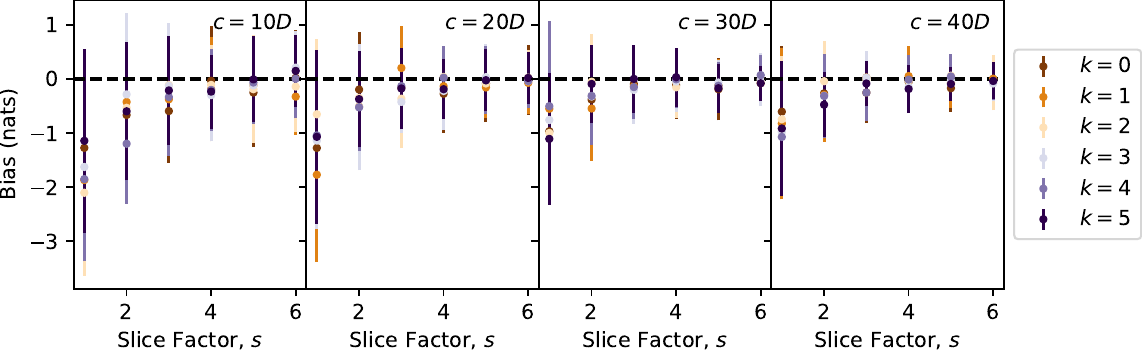}
    \caption{The log-evidence bias as a function of slice factor, $s$, for the $D=8$ model, coloured by the number of retained phantom samples, $k$ for different values of $c$. 
    The dashed line represents the target log-evidence distribution for standard nested sampling.
    Each point and error bar is ensemble averaged over 100 runs.}
    \label{fig:bias_vs_s_grid_per_c_8D}
\end{figure*}

We discuss here the results for the $D=8$ model, with the same results holding for the $D=16$ model which can be found in Appendix~\ref{app:16D_results}.

In Figure~\ref{fig:bias_vs_s_grid_per_c_8D} we can see the log-evidence bias as a function of slice factor $s$ for the $D=8$ model, for different numbers of parallel Markov-chains, $c$, and retained phantom points, $k$. 
We observe two well-known facts for standard nested sampling $k=0$ \citep[e.g.][]{2023StSur..17..169B}. 
Firstly, that runs with lower $s$ tend to be negatively biased due to autocorrelation of accepted samples, and as $s$ increases the log-evidence distribution asymptotes to a single distribution. 
Secondly, as $c$ increases the width of the log-evidence distribution becomes more narrow, i.e. evidence estimations become more precise.

In addition we observe for fixed values of $s$ and $c$ a variation between log-evidence distributions over values of $k$.
This implies that retaining phantom samples introduces an inaccuracy into the evidence calculation, as expected. 
However, as $s$ and $c$ increase we see this variation shrink.
Visually we note that the log-evidence statistics appear consistent for all runs with $s\ge 4$ and $c \ge 20 D$.
Since all runs within this region have consistent log-evidence bias statistics, we label this the \textit{consistent region}.
We note this region may be different for different models.

\begin{figure}
    \centering
    \includegraphics[width=\columnwidth]{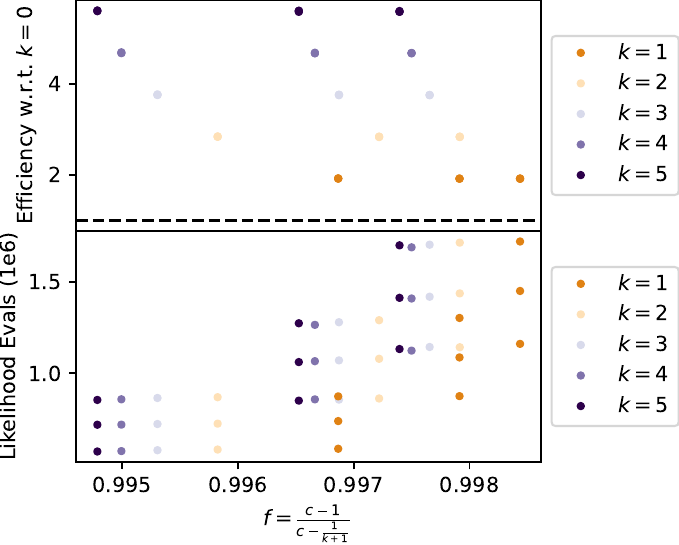}
    \caption{Top panel: The efficiency improvement vs $f$ for the $D=8$ model for runs in the consistent region, $s \ge 4$ and $c \ge 20 D$, relative to standard nested sampling, $k=0$. 
    Note, runs with the same $s$ have nearly identical efficiency improvement and so are indistinguishably plotted on top of each other.
    The dashed line is equality  with $k=0$. Bottom panel: The total number of likelihood evaluations vs $f$.
    Each point is ensemble averaged over 100 runs.}
    \label{fig:speedup_and_likelihood_vs_f_per_k_8D}
\end{figure}

We then explore the efficiency and cost per run in the consistent region.
Figure~\ref{fig:speedup_and_likelihood_vs_f_per_k_8D} shows relative efficiency improvements, relative to $k=0$, of all runs in the consistent region, against the fraction of totally random pairwise orderings, defined in Appendix~\ref{app:proof}.
As expected, as $k$ increases the sample efficiency goes up and the runs complete faster leading to an overall speed up relative to $k=0$.
An maximal efficiency improvement of 5 is observed for $k=5$, as expected.
In terms of number of likelihood evaluations, we observe that the lowest number of likelihood evaluations is observed for largest $k$, and minimal $s$ and $c$ within the consistent region.
Importantly, since all runs in the consistent region have consistent log-evidence distributions, we are at liberty to choose the run setting with the smallest required likelihood evaluations.
This result empirically supports the validity of the proposed method.

\section{Discussion}
\label{sec:discussion}

Taken together these results suggest that there is a natural sweet spot for accelerating NS, where minimal sufficiently large $s$ and $c$ should be selected and the highest possible $k>0$ number of phantom points retained.
Based on the results, we see no reason why higher values of $k$, than those explored here, could be used for even greater speed ups.
However, as $k \to s D - 1$ we expect autocorrelation to creep back into the samples, and the resulting evidence calculations to become biased. 
To explore this matter further we perform a quick ablation study to understand the impact of choosing larger values of $k$.
We run the method for the $D=8$ model with $s=6$ and $c=40 D$.
These high values of $s$ and $c$ are well within the consistency region, and ensures that accepted samples are independent. 
We then explore values of $k \in \{ 0, \ldots, s D - 1\}$, again taking ensemble averages over 100 random seeds.

\begin{figure}
    \centering
    \includegraphics[width=\columnwidth]{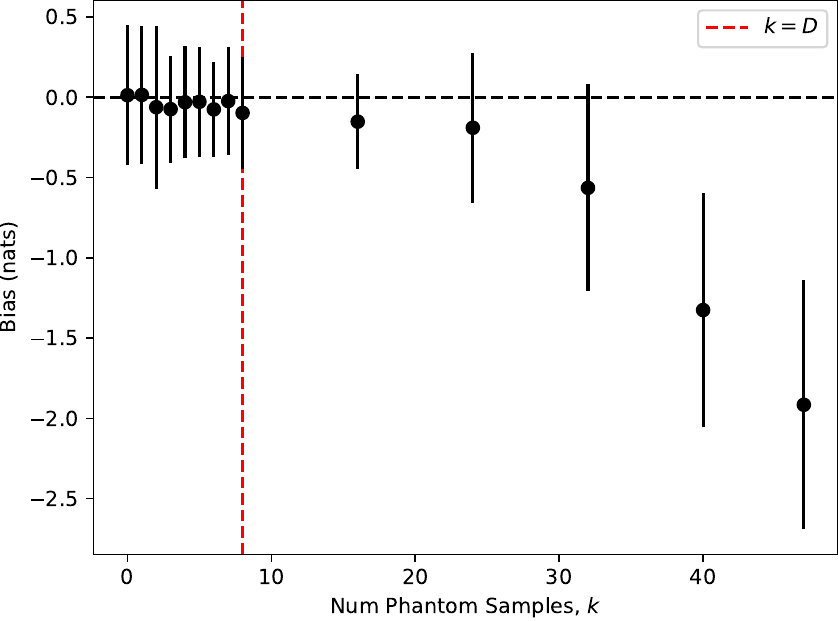}
    \caption{The bias for $s=6$ and $c=40 D$ as we vary $k$ from low values to its maximal value of $s D - 1$, for the $D=8$ model.
    At $k=0$ we have standard nested sampling.
    The red vertical line is $k= D$. 
    We observed that for $k \lesssim D$ the evidence statistics appear consistent, and for $k\gtrsim D$ they become biased as $k$ increases.
    Each point and error bar is ensemble averaged over 100 runs.}
    \label{fig:ablation_bias_vs_k_8D}
\end{figure}

Figure~\ref{fig:ablation_bias_vs_k_8D} shows the resulting log-evidence bias as a function $k$.
Indeed we observe that for low enough values of $k$ there is negligable impact on evidence calculations, however starting from around $k=D$ (the dashed red line) we see bias start to grow.
This confirms our expectation that there is a limit to the speed up attainable from this method. 
A tentative guess at the maximal number of retainable phantom points would be $k_{\rm max}=D$, which would lead to a maximal relative efficiency improvement of $D$, i.e. an efficiency improvement that grows with model dimension. 
It is unclear here if the bias increases due to breakage of the identicality or independence requirements of $\mathcal{S}$.
This finding should be verified with numerous other models, as well as larger values of $s$, before being trusted.
In particular, using larger values of $s$ may allow larger values of $k$ in a problem-dependent manner.

A question arises as to whether this method can be used for arbitrarily complex likelihood models.
If we look at the basic assumptions of the proposed method then there are no aspects that are dependent on the complexity of the likelihood. 
In fact all the hard work of dealing with difficult likelihoods has already been put into creating good Markov-chain samplers.
The only aspects which require model-dependent consideration are choosing minimally sufficiently larger $s$ and $c$.
This method takes advantage of a general property of all Markov-chain samplers, which is that after enough proposal samples all further samples are identically uniformly distributed within the likelihood constraint. 
It then takes advantage of a scaling relationship that asymptotically ameliorates the impact of inter-sample dependence during the shrinkage process.
So we do not have any reason to suppose this might not be effective for models of any complexity.

\section{Conclusions}
\label{sec:conclusion}

This paper introduces an enhancement to the nested sampling algorithm, particularly addressing the challenge of computational inefficiency in high-dimensional Bayesian evidence calculations. 
The crux of our method lies in the integration of `phantom points' -- previously disgarded due to their high autocorrelation -- into the evidence computation process. 
By ensuring these points remain asymptotically i.i.d. uniformly distributed we propose a method for them to be utilised effectively, contributing to the accuracy of the final evidence estimate while simultaneously boosting sampling efficiency.

Our theoretical argument, empirical evaluation, and ablation study have demonstrated that:
\begin{enumerate}
    \item sampling efficiency can be increased without degrading evidence accuracy;
    \item any ergodic Markov-chain likelihood-constrained sampler can be easily adapted to this method; and
    \item the method scales well with dimension.
\end{enumerate}

The reduction in likelihood evaluations is a critical advancement, particularly in fields like astrophysics, cosmology, and climate modelling, where computational resources are often a limiting factor.
Based on our discussion, a tentative maximal achievable speed up could be $D$, implying a potential efficiency that increases with model dimension.
However, this maximal speed is currently only weakly supported, and should be verified on a wider variety of models.
In the models we explored, we observed an efficiency improvement factor of 5.
Since the efficiency gains scale well with model dimension, our method helps opens the door to Bayesian analysis of larger data sets and models.

In addition, it should be noted that nested sampling has been applied outside the realm of Bayesian analysis, particularly in material sciences \citep[e.g.][]{2023arXiv230808509Y}, and it would be interesting to explore if the proposed method can be applied there.

\backmatter

\bmhead{Supplementary information}

All code and data required to reproduce the results of this study are provided in a git repository \url{https://www.github.com/joshuaalbert/jaxns}.

\bmhead{Acknowledgments}

JA would like to thank Do Kester and Romke Bontekoe for pointing out clarifications, and John Skilling for discussions around nested sampling, as well as Aleksandrina Skvortsova for helpful discussions around general compression.

\begin{appendices}

\section{Proving asymptotic i.i.d. uniformity of $\mathcal{S}$}
\label{app:proof}

\begin{figure}[b]
    \centering
    \def\svgwidth{\columnwidth}
\begingroup%
  \makeatletter%
  \providecommand\color[2][]{%
    \errmessage{(Inkscape) Color is used for the text in Inkscape, but the package 'color.sty' is not loaded}%
    \renewcommand\color[2][]{}%
  }%
  \providecommand\transparent[1]{%
    \errmessage{(Inkscape) Transparency is used (non-zero) for the text in Inkscape, but the package 'transparent.sty' is not loaded}%
    \renewcommand\transparent[1]{}%
  }%
  \providecommand\rotatebox[2]{#2}%
  \newcommand*\fsize{\dimexpr\f@size pt\relax}%
  \newcommand*\lineheight[1]{\fontsize{\fsize}{#1\fsize}\selectfont}%
  \ifx\svgwidth\undefined%
    \setlength{\unitlength}{273.04297606bp}%
    \ifx\svgscale\undefined%
      \relax%
    \else%
      \setlength{\unitlength}{\unitlength * \real{\svgscale}}%
    \fi%
  \else%
    \setlength{\unitlength}{\svgwidth}%
  \fi%
  \global\let\svgwidth\undefined%
  \global\let\svgscale\undefined%
  \makeatother%
  \begin{picture}(1,0.40426185)%
    \lineheight{1}%
    \setlength\tabcolsep{0pt}%
    \put(0,0){\includegraphics[width=\unitlength,page=1]{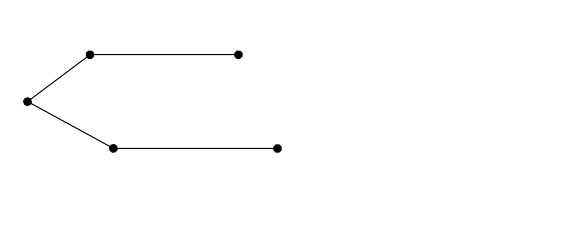}}%
    \put(-0.00332623,0.25515975){\color[rgb]{0,0,0}\makebox(0,0)[lt]{\lineheight{1.25}\smash{\begin{tabular}[t]{l}root\end{tabular}}}}%
    \put(0,0){\includegraphics[width=\unitlength,page=2]{sample_tree_multi.pdf}}%
    \put(0.25446005,0.0059729){\color[rgb]{0,0,0}\makebox(0,0)[lt]{\lineheight{1.25}\smash{\begin{tabular}[t]{l}$m=6$\end{tabular}}}}%
    \put(0.50167388,0.0059729){\color[rgb]{0,0,0}\makebox(0,0)[lt]{\lineheight{1.25}\smash{\begin{tabular}[t]{l}$\mathbb{E}[m]=c (k+1)$\end{tabular}}}}%
  \end{picture}%
\endgroup%

    \caption{A sample tree representation of $c=2$ parallel Markov-chains with phantom samples in green, and accepted samples in black. A vertical line at any likelihood value will cross on average $c (k+1)$ edges.}
    \label{fig:sample_tree_multi}
\end{figure}

Given $k$ phantom samples and $c$ parallel Markov-chains, we will prove that the likelihoods in a sorted ordering of $\mathcal{S}$ will asymptotically approach total randomness as $c$ increases.

We consider the situation where there are $c>1$ parallel Markov-chains, for which the sample tree is helpful for visualising.
Figure~\ref{fig:sample_tree_multi} shows two parallel Markov-chains. 
We retain $k$ phantom samples for every accepted sample. 
We assume that each accepted sample and its $k$ phantom samples are identically, but not independently, uniformly distributed within the same likelihood constraint.
Note, it can easily be seen that we have $m=c(k+1)$ crossed edges on average at every likelihood value along the sample tree.

Consider a likelihood set, $\mathcal{S}=\{\lambda_i\}_{i=1}^m$, and define the sorted ordering of likelihoods, $\sigma(\mathcal{S}) = (\sigma(1), \ldots, \sigma(m)\}$. 
Initially, suppose $k=0$ so that all likelihoods in $\mathcal{S}$ are identically \textit{and} independently uniformly distributed.
Thus, for any two selected indices $j,j'$ we have $\mathrm{Prob}(\sigma(j) < \sigma(j') ) = 1/2$, i.e. we have total randomness. 
Total randomness is sufficient to show that the probabilities for any sample, $j$, to have any sorted index, $i$, are all equal, i.e. $\mathrm{Prob}(\sigma(j) = i) = 1/m$. 

Now, suppose $k>0$ so that accepted samples and their phantom samples are no longer independently distributed, but still identically distributed. 
Recognise that accepted samples and their phantom samples, form cliques where -- due to their non-independence -- for all $j,j'$ in a clique $\mathrm{Prob}(\sigma(j) < \sigma(j')) \neq 1/2$.
However, we still have independence between cliques, so for $j,j'$ \textit{not} in the same clique $\mathrm{Prob}(\sigma(j) < \sigma(j')) = 1/2$, i.e. we have total randomness for inter-clique orderings.
We can think of this as \textit{relative} pairwise ordering being totally random between cliques, and not so for pairwise orderings within a clique.

Let us now count favourable outcomes of pairwise orderings which are totally random within $\mathcal{S}$.
Each likelihood is within a clique of size $k+1$ and forms an inter-clique pair with $m-(k+1)$ other likelihoods.
It follows that number of favourable outcomes is $m(m - (k+1))$, and the total number of pairwise orderings is $m(m-1)$.
Thus, using $m\approx c(k+1)$, the fraction of totally random pairwise orderings is,
\begin{align}
    f \triangleq& \frac{m - (k+1)}{m - 1}\\
    =&\frac{c - 1}{c - \frac{1}{k+1}}
\end{align}

Clearly, $f\to 1$ as $c \to \infty$.
Since $f=1$ is a sufficient condition for uniform likelihood orderings, therefore the likelihoods in a sorted ordering of $\mathcal{S}$ will asymptotically approach total randomness as $c$ increases. $\blacksquare$

\section{16-Dimensional Results}
\label{app:16D_results}

\begin{figure*}[t]
    \centering
    \includegraphics[width=2\columnwidth]{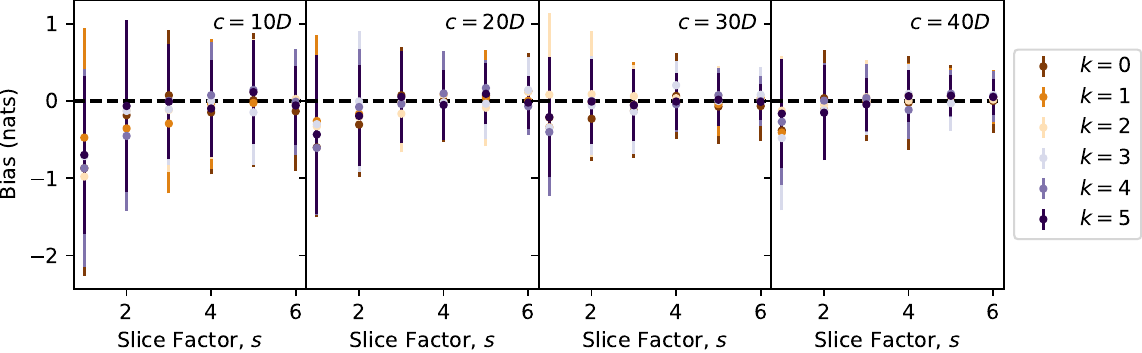}
    \caption{The log-evidence bias as a function of slice factor, $s$, for the $D=16$ model, coloured by the number of retained phantom samples, $k$ for different values of $c$. 
    The dashed line represents the target log-evidence distribution for standard nested sampling.
    Each point and error bar is ensemble averaged over 100 runs.}
    \label{fig:bias_vs_s_grid_per_c_16D}
\end{figure*}

Results for the $D=16$ model are provided here, with log-evidence bias as a function of $s$ for different values of $c$ and $k$ shown in Figure~\ref{fig:bias_vs_s_grid_per_c_16D}.
We visually see log-evidence distributions appear consistency for all $k$ for all runs in $s\ge 3$ and $c \ge 20 D$, which is the same consistency region observed for the $D=8$ model.

\end{appendices}

\bibliography{cite}

\end{document}